\documentclass[%
 reprint,
 superscriptaddress,
 amsmath,amssymb,
 aps,
prl,
floatfix]{revtex4-1}

\usepackage{graphicx}% Include figure files
\usepackage{dcolumn}% Align table columns on decimal point
\usepackage{bm}% bold math
\usepackage{hyperref}% add hypertext capabilities
\setcitestyle{numbers,square}
\usepackage{color}
\hypersetup{
	colorlinks = true,
	pdftitle = {},
	pdfauthor = {},
	pdfkeywords = {},
	linkcolor = blue,
	citecolor = blue,
	filecolor = black,
	urlcolor = magenta,
}

\begin{document}
\title{Imaging atomic-scale magnetism with energy-filtered differential phase contrast method}
% or "Differential phase contrast imaging at atomic scale using electron magnetic chiral dichroism"?
\author{Devendra Singh Negi}
\email{devendra@iitj.ac.in}
\affiliation{Department of Metallurgical and Materials Engineering, Indian Institute of Technology Jodhpur,  NH 62, Surpura Bypass Road, Karwar, Rajasthan 342030, India}

\author{Peter A. van Aken}
\affiliation{Stuttgart Center for Electron Microscopy, Max Planck Institute for Solid State Research, Heisenbergstr.1, 70569 Stuttgart, Germany}

\author{Jan Rusz}
\email{jan.rusz@physics.uu.se}
\affiliation{Department of Physics and Astronomy, Uppsala University, P.O. Box 516, 75120 Uppsala, Sweden}

\date{\today} 
 
%--- abstract section ---------------------------
\begin{abstract}
We propose differential phase contrast (DPC) imaging using energy-filtered electrons to image the magnetic properties of materials at the atomic scale. Compared to DPC measurements with elastic electrons, our simulations predict about two orders of magnitude higher relative magnetic signal intensities {\color{black}and} sensitivity to all three vector components of magnetization.%, and an enhanced robustness with respect to dynamical diffraction effects.
\end{abstract}
%------------------------------------------------
%\pacs{}
\maketitle
%----- Introduction starts here -----------------
%\section{Introduction}\label{sec:intro}
%-- first line for the introduction ------------

% §1: general paragraph about the need of nano-scale characterization of magnetic properties of materials and an overview of existing methods (NVE-magnetometry, STM-based methods, TEM-methods: holography, EMCD, Lorentz microscopy & DPC)
Nanostructured magnetic materials are used in a wide range of applications, spanning environmental, biomedical and industrial fields. These and future potential applications, including energy harvesting, magnetic sensors or refrigeration, drug delivery, data storage and others, explain the broad range and intensity of research initiatives in this area \cite{elgendi_magnetic_2018}. Characterization of magnetic materials at high spatial resolution, reaching nanometer or an atomic scale is an essential part in the development of devices. Several strategies exist, including surface-sensitive methods like nitrogen-vacancy center magnetometry \cite{welter_scanning_2022} and spin-polarized scanning tunneling microscopy \cite{wiesendanger_spin_2009} and bulk-sensitive methods using transmission electron microscopy (TEM), such as holographic methods \cite{kovacs_magnetic_2018,Tanigaki_electron_2024}, electron magnetic circular dichroism \cite{song_magnetic_2019}, 4D-STEM approach \cite{nguyen_angstrom_2023} and differential phase contrast imaging, which is in the focus of our work.

% §2: focus on DPC - history, achievements and limitations
Differential phase contrast scanning TEM (DPC-STEM), is an imaging technique in transmission electron microscopy, which is used to directly visualize the electric and magnetic fields \cite{dekkers_differential_1974,dpc_chapman}. DPC-STEM probes the elastically deflected incident electrons due to their interaction with the electric and magnetic fields in the material. It was successfully used to {\color{black}image} skyrmions \cite{dpc_skrymion} or electric fields in \emph{pn}-junctions \cite{dpc_pn_junction}. Using aberration-corrected STEM instruments, this technique has been pushed down to nanometer scale \cite{mcvitie_aberration_2015}. Aberration-corrected STEM enabled also the next step in the evolution of the method, namely, detection of microscopic electric fields around atoms \cite{dpc_shibata_first,dpc_EMF,dpc_qm_muller,hachtel_subangstrom_2018,dpc_deflection}. A detailed quantum mechanical treatment was presented in Ref.~\cite{muller_measurement_2017,dpc_segmented_detector} and advantages of pixelated detectors were analyzed in detail \cite{dpc_pixelated_detector1,dpc_pixelated_detector2}. A possibility to also detect microscopic magnetic fields in DPC-STEM has been discussed and computationally explored in Ref.~\cite{lubk_differential_2015,dpc_qm_rusz}. Reaching this milestone in experiments, microscopic magnetic fields in STEM-DPC were detected in 2022 in an antiferromagnet Fe$_2$O$_3$ \cite{dpc_AFM_nature} and utilized for mapping atomically thin antiferromagnetic domain walls in CuMnAs \cite{dpc_sharp_domain_science}.

Since its inception in 1974, the DPC-STEM went through tremendous improvements. Nevertheless, for atomic-scale studies, the DPC-STEM has also some limitations. Measurements are directly interpretable only when the sample is ultra-thin (typically well below 10~nm), otherwise the dynamical diffraction scrambles the images \cite{muller_measurement_2017,dpc_qm_rusz} and their interpretation then requires matching with simulations. Furthermore, the magnetic signal strength is typically several orders of magnitude weaker than the electric one \cite{dpc_qm_rusz,dpc_AFM_nature}. Finally, in STEM-DPC one has access only to the in-plane magnetic fields, which may require to switch off objective lens or to have a magnetic-field-free sample stage \cite{shibata_atomic_2019}.

% §3: what we did
In this {\color{black}Article} we show that the differential phase contrast method applied to energy-filtered diffraction patterns at core level edges can reveal magnetic information at the atomic scale and, compared to its elastic scattering counterpart \cite{dpc_AFM_nature}, it would {\color{black}provide the following} advantages: {\color{black}T}he relative strength of the magnetic signal improves by about two orders of magnitude{\color{black}, and,}
%Second, simulations indicate that this approach is more robust with respect to dynamical diffraction, in some case being applicable to samples of thickness of several tens of nanometers. Third, 
the magnetization component parallel to the incoming beam influences DPC images. {\color{black}W}e outline methods how to retrieve this information.

%----- figure 1 here ---------------------------
\begin{figure}
    \centering
    \includegraphics[width=\linewidth]{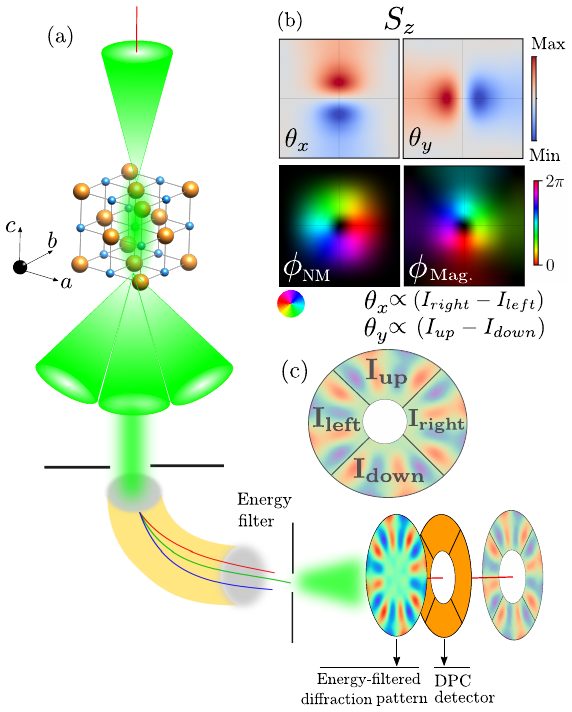}
    \caption{a) Schematic diagram representing the energy-filtered DPC-STEM method. The atomic size probe passes through the crystal in zone axis orientation and an energy filter with a slit is used to generate a diffraction pattern formed by electrons with energy-losses within a chosen energy interval. b) Example results of energy-filtered DPC-STEM assuming a magnetization parallel with the beam direction. The {\color{black}beam deflection angles $(\theta_x,\theta_y)$} and the {\color{black}corresponding azimuthal} angle of the {\color{black}beam deflection} $\phi$ are shown (see the color wheel for interpretation of colors as {\color{black}beam deflection} directions). The {\color{black}$\theta_x$ and $\theta_y$} are extracted using the indicated formulas {\color{black}(see also equations in the text below)}, where the intensities {\color{black}$I_\text{up},I_\text{down},I_\text{left},I_\text{right}$} come from four segments of the detector, as indicated in panel c).}
    \label{fig1}
\end{figure}

The scheme of the proposed experiment is shown in Fig.~\ref{fig1}a. A convergent electron beam of atomic size passes through the sample in zone-axis direction. Then, the electrons pass through an energy filter and only electrons with an energy-loss range corresponding to a chosen core-level edge would be allowed to pass. These electrons would form an energy-filtered diffraction pattern, which could be detected either by a pixelated detector or a segmented detector. Thus, the only difference from a standard DPC-STEM is the energy filtering of the scattered electrons.

\begin{figure}
    \centering
    \includegraphics[width=\linewidth]{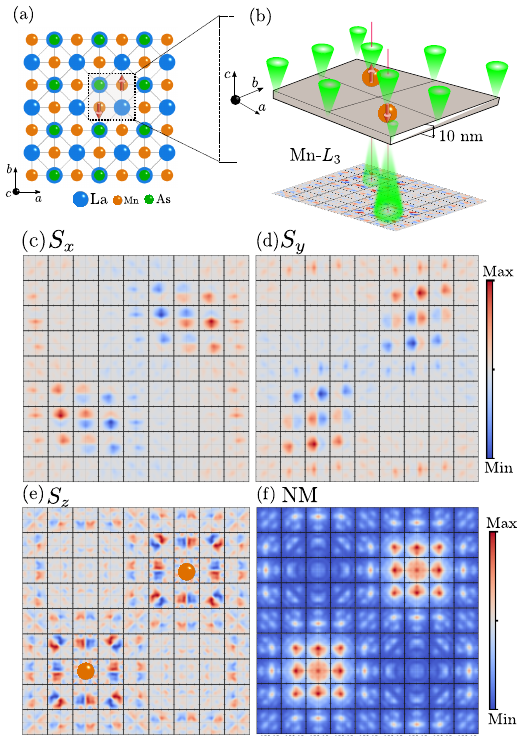}
    \caption{a) Projected crystal structure of LaMnAsO along the $c$-axis with indicated scanned region, as highlighted in panel b). Panels c) to f) show explicitly the Mn $L_3$-edge energy-filtered diffraction patterns for a grid on $9 \times 9$ beam positions. The panels show components of double-differential inelastic scattering cross-section due to $1\mu_B$ of magnetization in c) $x$-direction, d) $y$-direction, e) $z$-direction and f) a non-magnetic component normalized per hole in the 3$d$-shell.}
    \label{fig2}
\end{figure}

As a model system, we have chosen LaMnAsO crystal,  {\color{black}{which is antiferromagnetic in nature, thus offering a platform for exploring the feasibility of imaging antiferromagnetism by the proposed method}}. It has a tetragonal crystal structure with $a=4.12$~\AA{} and $c=9.03$~\AA{} and an antiferromagnetic arrangement of magnetic moments of Mn atoms is such that along the $c$-axis the moments are always parallel within an atomic column, while neighboring atomic columns have opposite signs of magnetic moments. All magnetic moments are parallel to the $c$-axis \cite{emery_giant_2010}. Arrangement of atomic columns can be seen in Fig.~\ref{fig2}a, assuming an electron beam parallel to the crystallographic $c$-axis.

We have calculated inelastic scattering intensities by using MATS.V2 software, which combines the multislice and Bloch-waves approaches for calculating the double differential scattering cross-section \cite{MATS_V2}. {\color{black}This is in contrast with simulations of magnetic DPC-STEM in the elastic regime \cite{dpc_qm_rusz} utilizing the Pauli multislice method \cite{edstrom_elastic_2016}.} The {\color{black}MATS.v2 method neglects the weak magnetic effects in elastic propagation, but it} offers conveniently to calculate {\color{black}inelastic scattering} intensities integrated over an energy range of a selected core-level edge and to separate the intensities into non-magnetic and magnetic contributions, the latter three also separated into contributions {\color{black}due to} magnetic moments along individual $x,y,z$-directions. {\color{black}The magnetic effects in the inelastic regime (EMCD) reach relative strengths of the order of 10\%, while in the elastic regime the relative strenghts of $10^{-4}$ are typically observed \cite{loudon_antiferromagnetism_2012,dpc_AFM_nature}.} The convergence parameter in MATS.V2 was set to $\mathrm{5\times10^{-7}}$. The acceleration voltage in TEM was set to 100~kV and the convergence semi-angle of the incoming beam to {\color{black}20}~mrad. Calculations have covered scattering angles within the range from -25~mrad to 25~mrad in both $\theta_x,\theta_y$ directions, respectively{\color{black}, with a step of 2~mrad}. 

{\color{black}We have first performed simulations in the $(001)$ zone axis orientation, i.e., with beam direction parallel with the $c$-axis of LaMnAsO, which is also parallel with the magnetic moment directions.} An array of {\color{black}$8 \times 8$} beam positions has been considered, spanning {\color{black}the area of the entire unit cell. Using the crystal periodicity, this has been extended to $9 \times 9$ beam positions for visualization purposes in Fig.~\ref{fig2} to include all borders around the unit cell.}
%one quarter of the unit-cell area in $a,b$-plane, centered around a selected Mn atomic column, see panels a) and b) of Fig.~\ref{fig2}. 
Energy-filtered diffraction patterns were calculated for the Mn-$L_3$ edge (approximate energy-loss range of 640--650~eV). See Ref.~\cite{negi_probing_2018} for more details about the computational method. For the purpose of model considerations, in simulations we will assume not only the experimentally observed magnetic moment direction parallel to $c$-axis (or $z$-direction), but we will also present results for hypothetical magnetic moment directions along the $a$- and $b$-axes ($x$- and $y$-directions, respectively).

Figure~\ref{fig2}, panels c) to f) summarize the results of our calculations for a 10~nm thick LaMnAsO sample. They show four sets of $9 \times 9$ energy-filtered diffraction patterns, where each set shows only a particular contribution to the differential scattering cross-section: panels c), d) and e) show intensities due to magnetism, assuming the magnetic moment oriented along the $x$-, $y$- and $z$-direction, respectively, and panel f) represents the non-magnetic intensity. When the magnetic moments lie in-plane parallel to the $x$-direction, characteristic up/down features appear in the diffraction patterns, together with an overall negative or positive background intensity appearing in the neighborhood of the atomic column at scan positions on the left or right side, respectively, for a viewing direction parallel to the magnetic moment. The same holds true for the magnetic moments pointing in $y$-direction, see Fig.~\ref{fig2}d being identical as Fig.~\ref{fig2}c, only rotated 90~degrees anticlockwise.

For magnetic moments along the $z$-direction, the situation is more complicated and rich patterns appear depending on the beam position relative to the Mn atomic column. Nevertheless, intensities are non-zero, opening for a possibility to detect also magnetization along the $z$-direction. {\color{black} In the context of DPC imaging, one typically focuses on magnetization components perpendicular to the beam direction. However, in the EELS regime, magnetization along the beam direction gives rise to modifications of scattering cross-section due to electron magnetic circular dichroism (EMCD;  \cite{schattschneider_detection_2006}). It has been discussed previously that shifting the beam results in an asymmetric redistribution of the EMCD signal in a direction perpendicular to the beam shift \cite{Schattschneider_mapping_2012,emcd_atomic_plane_rusz,rusz_localization_2017}. This gives rise to an exciting possibility to map magnetic moments parallel to the beam direction with the DPC-like methods in the EELS regime. The expected beam deflection vectors due to magnetism should then be perpendicular to the direction of the beam shift with respect to the atomic column. This contrasts with the deflection vectors in the elastic scattering, which point towards the atomic column\cite{dpc_shibata_first,muller_measurement_2017}.}

Finally, the tableau of the non-magnetic component of the diffraction patterns presented in Fig.~\ref{fig2}f shows features that are qualitatively very similar to observations made in the {\color{black}elastic} regime, see for example \cite{muller_measurement_2017,hachtel_subangstrom_2018}. A circular ronchigram (however, here consisting of inelastically scattered electrons only) is being deformed as the beam is shifted around the atomic column. Thus, it can be expected that post-processing such diffraction patterns assuming a 4-segment detector or via evaluation of center of the mass vectors will lead to doughnut-shaped features, centered at the atomic sites, as in the elastic DPC-STEM.

%We explored DPC-EMCD in three different test bed materials, $\mathrm{LaMnAsO}$, $\mathrm{Fe}$-bcc[Supplementary], $\mathrm{FePt}$[Ssupplmentary].

%-----------------------------------------------
%\section{Results and Discussions}
%-----------------------------------------------

%--- figure 3 here -----------------------------
\begin{figure}
    \centering
    \includegraphics[width=\linewidth]{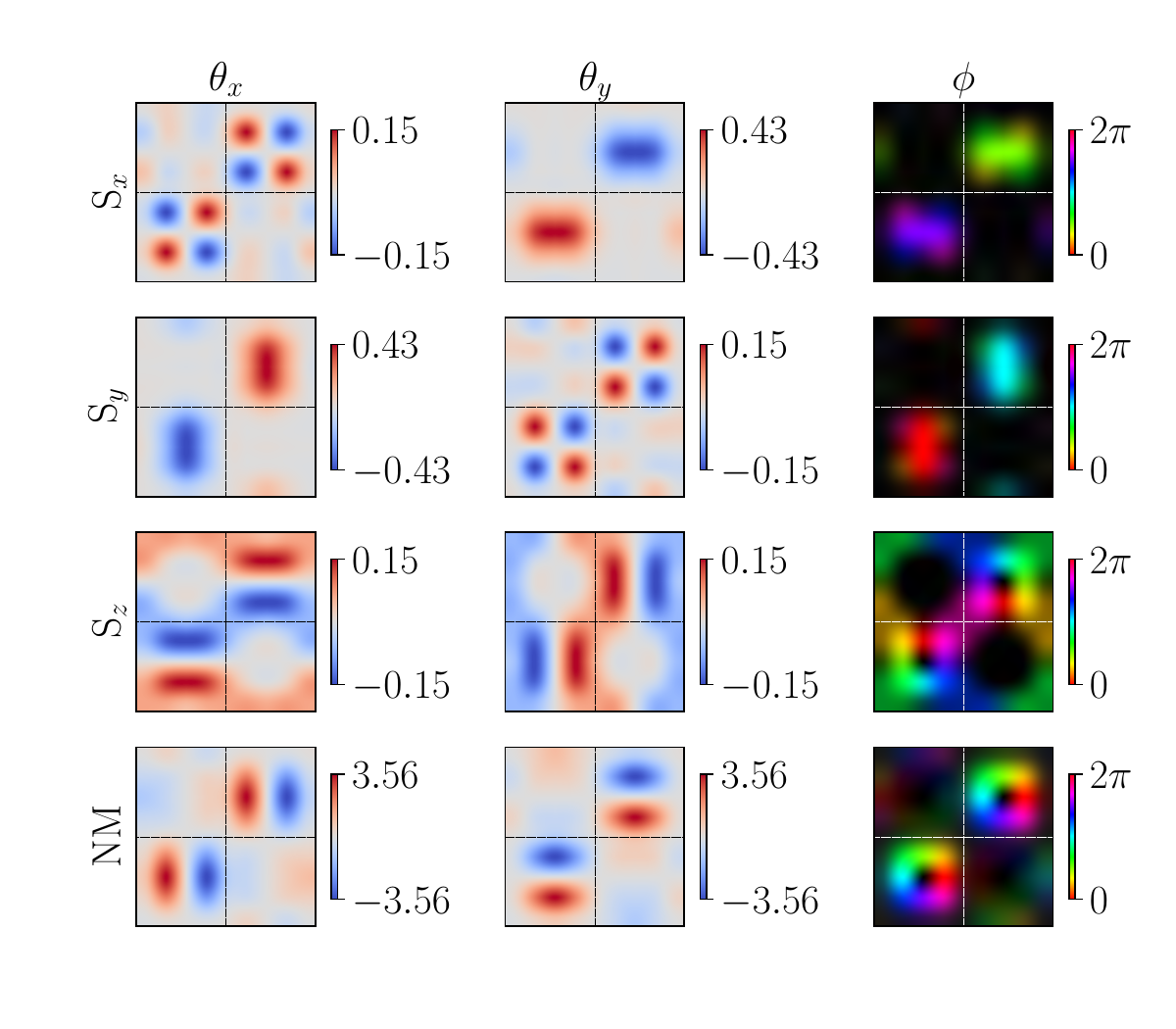}
    \caption{Energy-filtered DPC-STEM results for LaMnAsO with 10~nm sample thickness. Rows from top to bottom show electron {\color{black}beam deflection angles} due to magnetism, assuming a magnetic moment orientation along the $x$-, $y$- or $z$-direction, and the non-magnetic intensity. Columns from left to right show the {\color{black}deflection angles $\theta_x, \theta_y$} (see Fig.~\ref{fig1} {\color{black}and text below} for their definition), and magnitude and {\color{black}direction of the $(\theta_x, \theta_y)$} vector. The color wheel {\color{black}in Fig.~\ref{fig1}b} helps to interpret {\color{black}directions of beam displacement vectors.}}
    \label{fig3}
\end{figure}

{\color{black}For all of the beam positions, we have calculated the beam deflection vectors $(\theta_x, \theta_y)$. A four-segment detector has been assumed with inner and outer collection semi-angles set to 6~mrad and 24~mrad, respectively. The detector segments, Fig.~\ref{fig1}c, have their center of mass at angles $\vartheta=\pm 16.4$~mrad in a horizontal or vertical direction, respectively. The beam deflection angles are then obtained by
\begin{eqnarray*}
    \theta_x & = & \frac{I_\text{right}-I_\text{left}}{I^\text{NM}_\text{tot}} \vartheta \\
    \theta_y & = & \frac{I_\text{up}-I_\text{down}}{I^\text{NM}_\text{tot}} \vartheta
\end{eqnarray*}
where $I^\text{NM}_\text{tot}$ is a sum of non-magnetic intensities from all four detector segments. Note that this value depends on the beam position. In the evaluation of the magnetic and non-magnetic intensities \cite{emcd_single_atom_negi} we have assumed a spin magnetic moment of $2.43\mu_B$ per Mn atom and 5 holes in its 3$d$ shell, based on density functional theory calculations in Ref.~\onlinecite{Idrobo_detecting_2016}.
}

Results of this analysis are shown in Fig.~\ref{fig3}. 
%A four-segment detector has been assumed with inner and outer collection semi-angles set to 6~mrad and 24~mrad, respectively. 
%By following the usual procedure for four-segment detectors, we have extracted the intensity difference between the right and left quadrant as the $x$-component, and the intensity difference between the upper and lower quadrant as the $y$-component representing the shift of the beam intensity due to scattering. 
Processing all the diffraction patterns from Fig.~\ref{fig2}, we have obtained real-space images of $x$- and $y$-components of the {\color{black}beam deflection angles (in mrad)}, shown in the first and second column of Fig.~\ref{fig3}, respectively. %The overall amplitude of the shift vector is shown in the third column and the % 
%The phase (arc-tangent of the ratio of the $y$- and $x$-components of the shift vector) is shown in the {\color{black}third} column.
{\color{black}The direction [or phase $\phi = \arctan (\theta_y/\theta_x)$] and size of the beam deflection vectors are encoded in the hue and brightness of the colors, see the color wheel in Fig.~\ref{fig1}b.}

We start the discussion of the results for the non-magnetic intensities, shown in the bottom row. Note that we have obtained a volcano-shaped intensity profile and a vortex-like phase distribution, exactly as in the DPC-STEM in the elastic regime. The directions of the {\color{black}deflection} vectors, encoded in the phase image, point towards the center of the atomic column, again in close analogy with elastic DPC-STEM at atomic resolution. This suggests that atomic-scale DPC-STEM can also be realized by using energy-filtered inelastically scattered electrons, in an element-selective way. The price paid would be, of course, the significantly reduced electron counts. %Note the range of color bars in the first three columns, which have been obtained under an assumption of 100~pA beam current and 50~ms dwell time during acquisition \cite{negi_prospect_2019}.

The top row of Fig.~\ref{fig3} shows the DPC analysis due to magnetization along the $x$-direction. The horizontal {\color{black}deflection} of the beam{\color{black}, $\theta_x$,} shows a relatively complicated pattern with alternating positive and negative {\color{black}deflections} in the four quadrants surrounding the Mn atomic column. However, the amplitude of these {\color{black}deflections} is relatively small, almost two orders of magnitude lower than the {\color{black}deflections} due to non-magnetic scattering. On the other hand, the {\color{black}deflection} in the $y$-direction{\color{black}, $\theta_y$,} is of significantly larger magnitude and it is of the same sign {\color{black}within a region surrounding individual Mn atomic columns}. This is clearly seen in the phase image (the {\color{black}third} column), indicating that the beam {\color{black}deflection} due to interaction with the in-plane magnetization has a uniform direction in the vicinity of an atomic column. %Strikingly, in the closest vicinity of the atomic column, the shift due to the magnetic signal is above 10\% of the non-magnetic shift. Compared to the elastic regime, where the magnetic component of DPC reaches at best a fraction of a percent, in the inelastic regime we predict about two orders of magnitude higher relative strength of the magnetic shift.
The beam shift in response to magnetization along the $y$-direction has an analogical behavior.

Intriguing is the case of the beam response to the magnetization in the $z$-direction. In the elastic regime, the magnetization component parallel to the incoming beam is invisible. However, in the inelastic regime, we observe non-negligible beam {\color{black}deflections}, which mimic the behavior due to a non-magnetic component. The intensity profile of the {\color{black}deflections} has a similar doughnut shape around the atomic column, albeit with reduced intensity in the range of about 5\% (yet still significant). The fundamental difference is, however, in the directions of these {\color{black}deflections}. While the non-magnetic {\color{black}deflections} bend the beam towards the atomic column, the interaction with magnetization parallel to the beam {\color{black}deflects} the beam to a direction perpendicular with respect to the direction towards the atomic column{\color{black}, in line with the discussion above about the effect of beam shift on EMCD distribution}. See the phase image, column {\color{black}3}, where the color wheel is rotated by 90~degrees with respect to the one for non-magnetic component. This qualitatively different behavior opens for opportunities to isolate it from the non-magnetic beam {\color{black}deflection}. %In summary, our simulations suggest that the DPC-STEM realized on core-level edges is sensitive to all three components of the magnetization, including the one parallel to the incoming beam.
{\color{black}Note also that the rotation is clockwise for one of the Mn atomic columns and anti-clockwise for the other one, reflecting the opposite signs of the magnetic moments in these two atomic columns.}

{\color{black}It is striking that the beam deflection angles in the EELS regime are significantly larger than those observed in atomic resolution DPC-STEM images in the elastic regime. Kohno et al.~\cite{dpc_AFM_nature} report non-magnetic deflections on the order of 0.5~mrad and the average magnetic displacements on the order of 1~$\mu$rad. In our simulations, the beam displacements due to the non-magnetic component of the scattering cross-section are about an order of magnitude higher, reaching approximately 5~mrad. However, the magnetic component of scattering cross-section causes displacements of over 0.1~mrad, i.e., two orders of magnitude larger than in the elastic regime. This suggests that as long as the fractional intensity of the core-level edges is above $(1/100)^2$ of the zero loss peak, the signal to noise ratio of magnetic DPC-STEM has the potential to be higher at the core-level edges than in the elastic regime.}

%--- figure 4 here -----------------------------
\begin{figure}
    \centering
    \includegraphics[width=\linewidth]{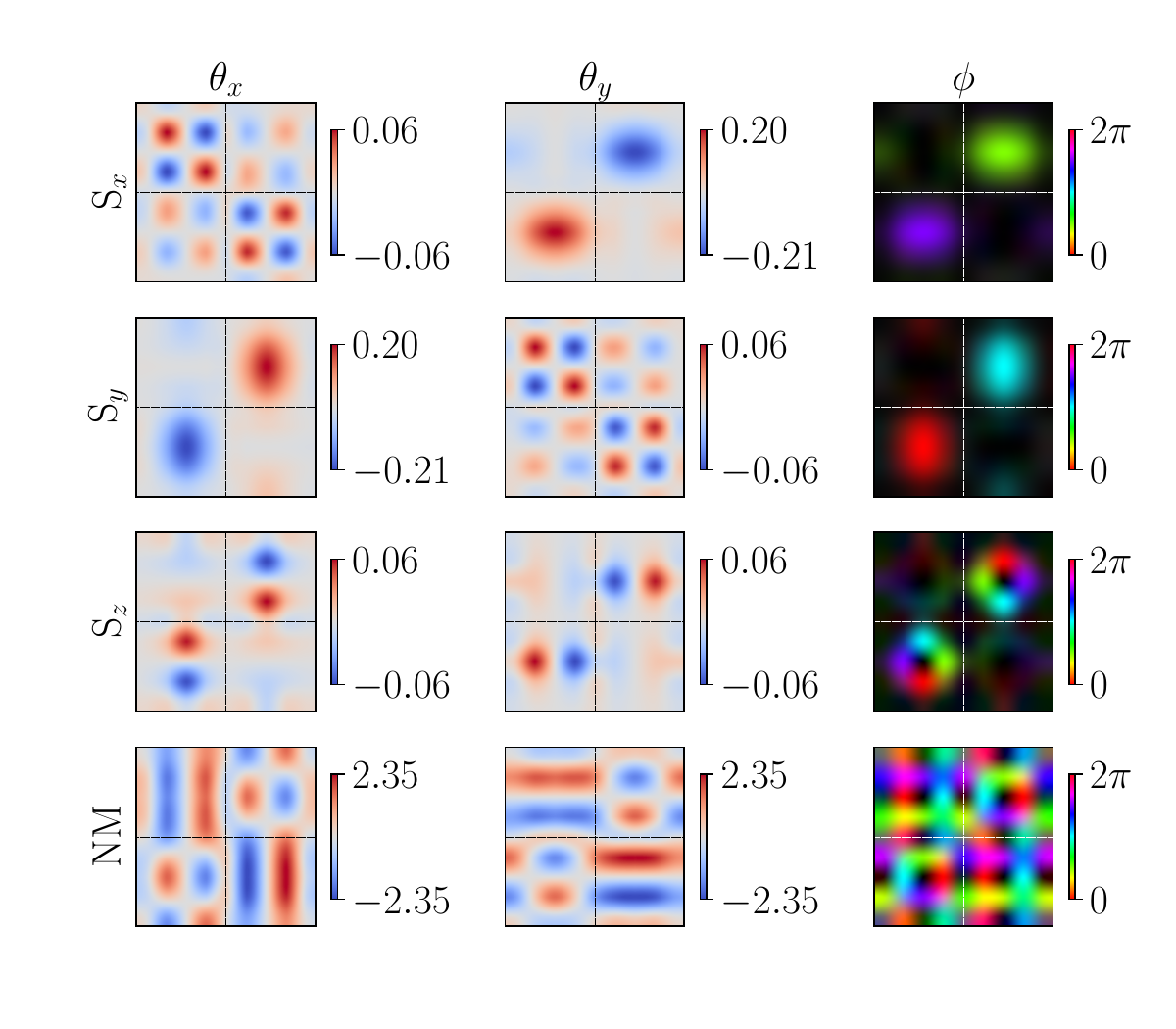}
    \caption{The same as Fig.~\ref{fig3} for a sample thickness of 20~nm.}
    \label{fig4}
\end{figure}

{\color{black}We have also performed simulations for a thicker sample.} Figure~\ref{fig4} shows results analogical to Fig.~\ref{fig3}, calculated for a LaMnAsO sample with a thickness of {\color{black}2}0~nm. At this sample thickness, the DPC-STEM images in the elastic regime are expected to be strongly influenced by dynamical diffraction and in a consequence they are rarely interpretable even qualitatively \cite{dpc_qm_muller}. 
{\color{black}We also observe a strong distortion of the beam displacement due to a non-magnetic component, which doesn't even allow to determine the position of Mn atomic columns. On the other hand, the displacements due to magnetic components to a large degree retain their shapes. This curious observation has likely a limited use, because the magnetic displacements will always be combined with the non-magnetic ones. Nevertheless, it might be possible to devise methods that would detect qualitative differences in displacement maps due to antiferromagnetism in the vicinity of otherwise equivalent Mn atomic columns.}
%Here, at core-level edges we see qualitatively the same results at 30~nm as at 10~nm. Absolute magnitudes of shifts have been modified by the propagation through a thick sample, nevertheless, the shape of the images remains unchanged. Supplementary figures~S1-S1{\color{black}3} show additional results for LaMnAsO as well as for bcc iron and FePt. Dynamical diffraction effects are observed in some cases, yet the simulations of energy-filtered DPC-STEM imaging indicate a higher robustness with respect to sample thickness, compared to its elastic counterpart.\newline

{\color{black}As it was indicated above, the phase map offers a pathway to image the nature of magnetic moment alignment in the materials. The phase maps of the magnetic signal are precisely inverted and represent the antiferromagnetic alignment of Mn atoms in LaMnAsO. However, the phase map of the nonmagnetic signal shows the same profile for both Mn atomic columns. Although, the phase map rotates as a function of the thickness of the specimen (compare $M_z$ maps in Fig.~\ref{fig3} with Fig.~\ref{fig4}), yet such nature remains invariant. Therefore, the inelastic DPC method could be adopted for direct imaging the nature of magnetism at the atomic scale.}

%-----------------------------------------------
%--- figure 5 here -----------------------------
\begin{figure}
    \centering
    \includegraphics[width=\linewidth]{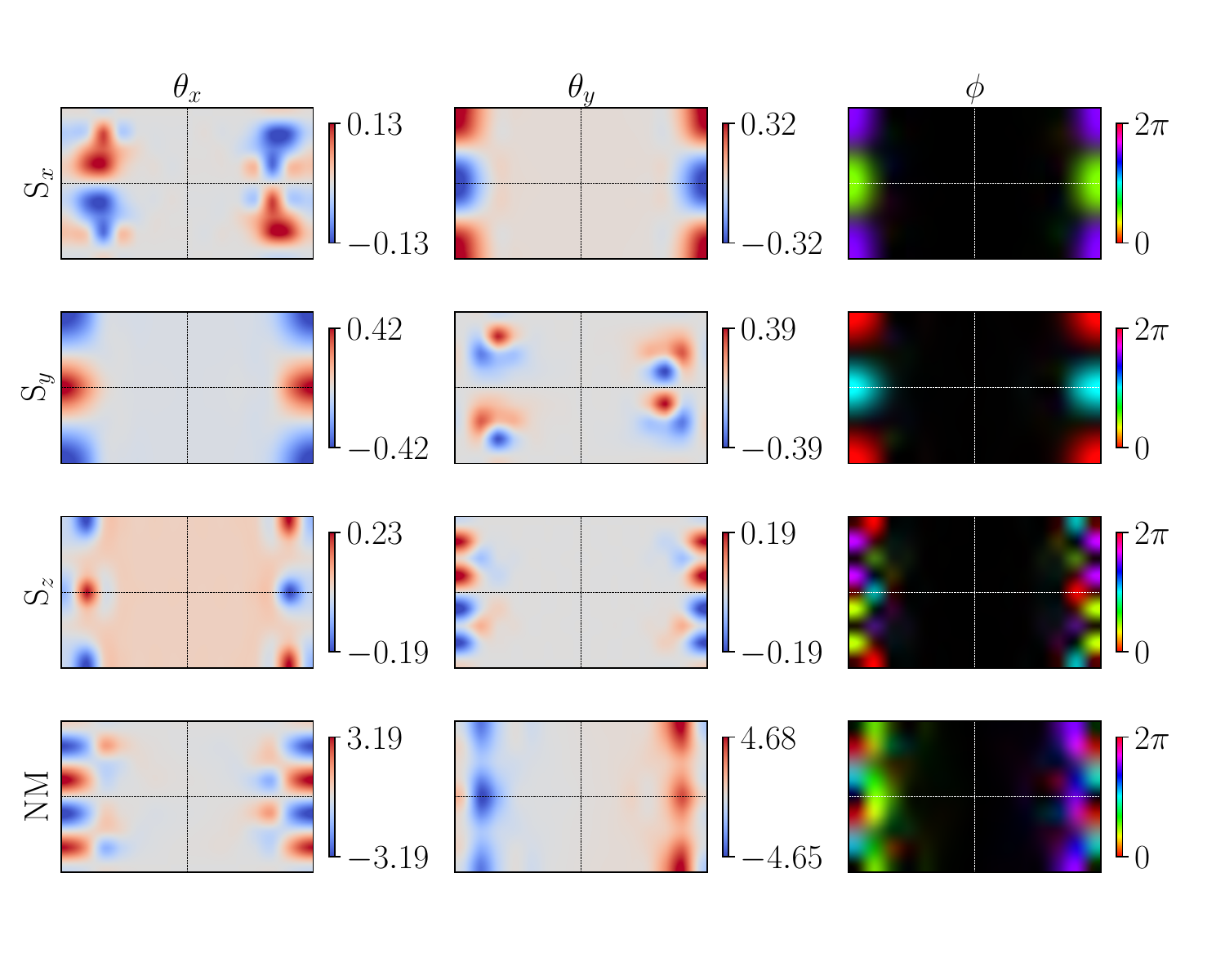}
    \caption{{\color{black}Energy-filtered DPC-STEM results for LaMnAsO
with 10 nm sample thickness with an electron beam incoming along the (100) zone axis. For other details see Fig.~\ref{fig3}.}}
    \label{fig5}
\end{figure}
%-----------------------------------------------
%\section{Conclusions}

{\color{black}To further reinforce our results, we have performed simulations for LaMnAsO with the incident beam parallel to the $[100]$ zone axis. This provides a realistic scenario to observe magnetic moments that are perpendicular to the beam direction, namely along the $x$-axis, corresponding to crystallographic $c$-axis. Results are summarized in Fig.~\ref{fig5}, along with other (hypothetical) directions of magnetic moments. We have used the same computational parameters, except for the grid of beam positions in STEM calculation being set to $8 \times 14$, reflecting the shape of the unit cell, while allowing for a larger scan step along the $c$-axis.

Along this beam direction, the projected distances of the Mn atomic columns are only about 2~\AA{}, leading to an overlap of the doughnut-shaped features around the atomic columns in the non-magnetic component of the beam displacement. Also, there is a large ``empty space'', highlighting the element-selectivity of the inelastic DPC-STEM method, since the Mn atoms are all located within one plane. Yet, all the qualitative features observed in the $(001)$ zone axis remain valid, both in terms of the directions and relative magnitudes of the beam deflections.}

In summary, on the basis of detailed simulations, we propose DPC-STEM experiments on energy-filtered diffraction patterns as a highly promising method for probing magnetism at the atomic-scale. Successful realization of such experiments would have a great impact on the development of functional magnetic nanostructures.\\
%-- acknowledgement here ------------------------
\begin{acknowledgements} 
This project has received funding from Science and Education Research Board (SERB), India under grant SRG/2022/000825. The SEED grant No: I/SEED/PRJ/DSN/AB/20220044 from IIT Jodhpur is acknowledged. {\color{black}J.R.\ acknowledges the support of the Swedish Research Council (grant no.\ 2021-03848), Olle Engkvist's foundation (grant no.\ 214-0331), and Knut and Alice Wallenbergs' foundation (grant no.\ 2022.0079). The simulations were enabled by resources provided by the National Academic Infrastructure for Supercomputing in Sweden (NAISS) at NSC Centre partially funded by the Swedish Research Council through grant agreement no. 2022-06725.}
\end{acknowledgements} 
%-----------------------------------------------
%\bibliographystyle{}

%\bibliography{ref}

%===============================================	
\end{document}